\documentclass[12pt,a4paper]{article}
\usepackage{amsmath}
\usepackage{graphicx}
\usepackage{ulem}
\usepackage{times}
\usepackage{cite}
\begin{document}
	\begin{titlepage}
		\title{Direct--channel   { option} of  the  {forward slope increase}}
		\author{ S.M. Troshin, N.E. Tyurin\\[1ex]
			\small   NRC ``Kurchatov Institute''-- IHEP\\
			\small   Protvino, 142281, Russian Federation}
		\normalsize
		\date{}
		\maketitle

		\begin{abstract}
		The LHC data on the elastic scattering indicate that the { forward slope increase} is not consistent with  the  { contributions of the simple Regge poles only with the } { linear } Regge trajectories. The  dynamics might be associated with  unitarization
		 in the direct channel of reaction. { We discuss the problems of the Regge model and provide a respective illustration of the direct--channel option.}
		\end{abstract}
	\end{titlepage}

\section{Introduction { and} problems with  { forward slope increase} revealed at the LHC}
The recent LHC measurements of the differential cross--section of $pp$--scattering have revealed new trends in the  { forward slope increase}. This is a speed up of   { this} steady increase \cite{totem, atlas, tsr}.
A linear logarithmic shrinkage of the diffraction cone { at $t=0$} and its relation 
to  the Regge trajectory parameter $\alpha'(0)$\footnote{ The Regge model predicts the scattering amplitude to have the form $F(s,t)\sim s^{\alpha(t)}$. The standard assumption of the  model corresponds to the choice of the linear trajectories with $t$--dependence  in the form $\alpha(t)=\alpha(0)+\alpha'(0)t$. } was  an evident success  of the Regge model  application   to the relativistic hadron scattering  (cf. e.g.\cite{m94}).  All the secondary Regge trajectories have different values of  intercepts $\alpha_{R{_i}}(0)$ ($\alpha_{R_i}(0)-1<0$) but  positive and approximately the same values of  $\alpha_{R_i}'(0)\simeq 0.9$ $(GeV/c)^{-2}$, i.e. the trajectories are linear and almost parallel (see for a recent review \cite{sch}).  The Pomeron trajectory has been introduced to reconcile the Regge model with  the experimental data on the total cross--sections. Currently this trajectory has an intercept $\alpha_P(0)-1>0$ and positive slope $\alpha_{P}'(0)\simeq 0.25$ $(GeV/c)^{-2}$ or even  less \cite{dl},  i.e. it is significantly lower than $\alpha'_R(0)$. Since $\alpha_P(0)-1>0$, the respective model amplitude needs to be unitarized. { Absence of unitarization would lead to the conflict of the partial amplitude magnitude at high energies with the unitarity limit.}

The main  assumption of the Regge model is an unjustified replacement of the whole amplitude by its asymptotic form in the cross channel. This is reformulated as the statement that  asymptotic amplitude analytically continued from the cross--channel   will give asymptotics of the amplitude in direct channel of the reaction.

 It appeared that  the actual rate of the diffraction cone  parameter
  \begin{equation}\label{bs0}
 B(s) \equiv \frac{d}{dt}\ln \frac{d\sigma}{dt}|_{t=0}
 \end{equation}
 growth exceeds the linear logarithmic extrapolation to the LHC energy range. In Eq. (\ref{bs0}), $d\sigma/dt $ is a differential cross--section of the proton--proton elastic scattering.
 To bring  the Regge model closer to the experiment on  the diffraction cone  { slope}   growth rate  it has been suggested that  the slope of the Pomeron trajectory $\alpha'_P(0)$ is an energy--dependent  function (cf. \cite{rysk}). Such a dependence    can reflect the negative contribution of  absorption correction due to increasing contribution of multi--Pomeron exchanges. It might  be due to Odderon presence also \cite{basar},  while complication of the Regge model in the form of Dipole Pomeron itself does not help \cite{ben}. 
 
 Regarding the possible form of transition from a linear logarithmic dependence 
 $$
 B(s)=B_0+2\alpha'_P(0)\ln s
 $$
  of  the Regge model { with poles } only { in the region of the Pomeron dominance} to the asymptotic $\ln^2 s$ dependence \cite{bas} the following observation should be mentioned here.
 The use of function 
 \begin{equation}\label{pol}
  B(s)=B_0+A\ln s+C\ln^2 s
  \end{equation}
  to fit the data  provides a negative or zero  value for the factor $A$ \cite{rysk},
 \[
 A \leq 0,
 \]
  and therefore  leads to the problems with its consideration as  the Pomeron trajectory slope  $\alpha'_P(0)${ inherited from the Regge--pole model}. The use of nonanalytic functions for  description of $B(s)$ (hints for a possible threshold energy dependence of $B(s)$ are given in \cite{tsr}) can preserve a positive factor in front of $\ln s$ term, but requires  extra justifications for a nonanalytical form of the scattering amplitude.
 
Even putting aside these fits, one should note that there is another problem regarding   the Regge model predictions in the  {whole} energy range. Namely, many papers suggest that the secondary Regge trajectories give a negligible contribution to the elastic  amplitude  and  the Pomeron  contribution is   the only significant one at the LHC energies. Moreover, it has been claimed    that a contribution of the secondary trajectories can already be neglected  at lower energies of $\sqrt{s}\sim 100$ $GeV$  \cite{rysk}. 
 
 The latter statement on the Pomeron dominance proceeds from the analysis of  the total cross--sections data. 
 However, since the Pomeron trajectory has a significantly lower slope $\alpha'_P(0)$ compared to $\alpha'_R(0)$ and conclusion on  vanishing contributions of  the secondary trajectories inevitably leads to associated prediction of slowdown  of $B(s)$ increase
in the  energy range where the Pomeron starts to dominate.  Thus, the Regge--pole model predicts break of $B(s)$ due to vanishing secondary poles' contributions and transition to the regime 
$$
B(s)= B_0+2\alpha'_R(0)\ln s
$$
at low energies\footnote{Note that $\alpha'_R(0)/\alpha'_P(0)\simeq 4$.}. Schematically, the predicted regimes in the energy dependence of $B(s)$ can be depicted at Fig. 1.
\begin{figure}[hbt]
		\resizebox{13cm}{!}{\includegraphics*{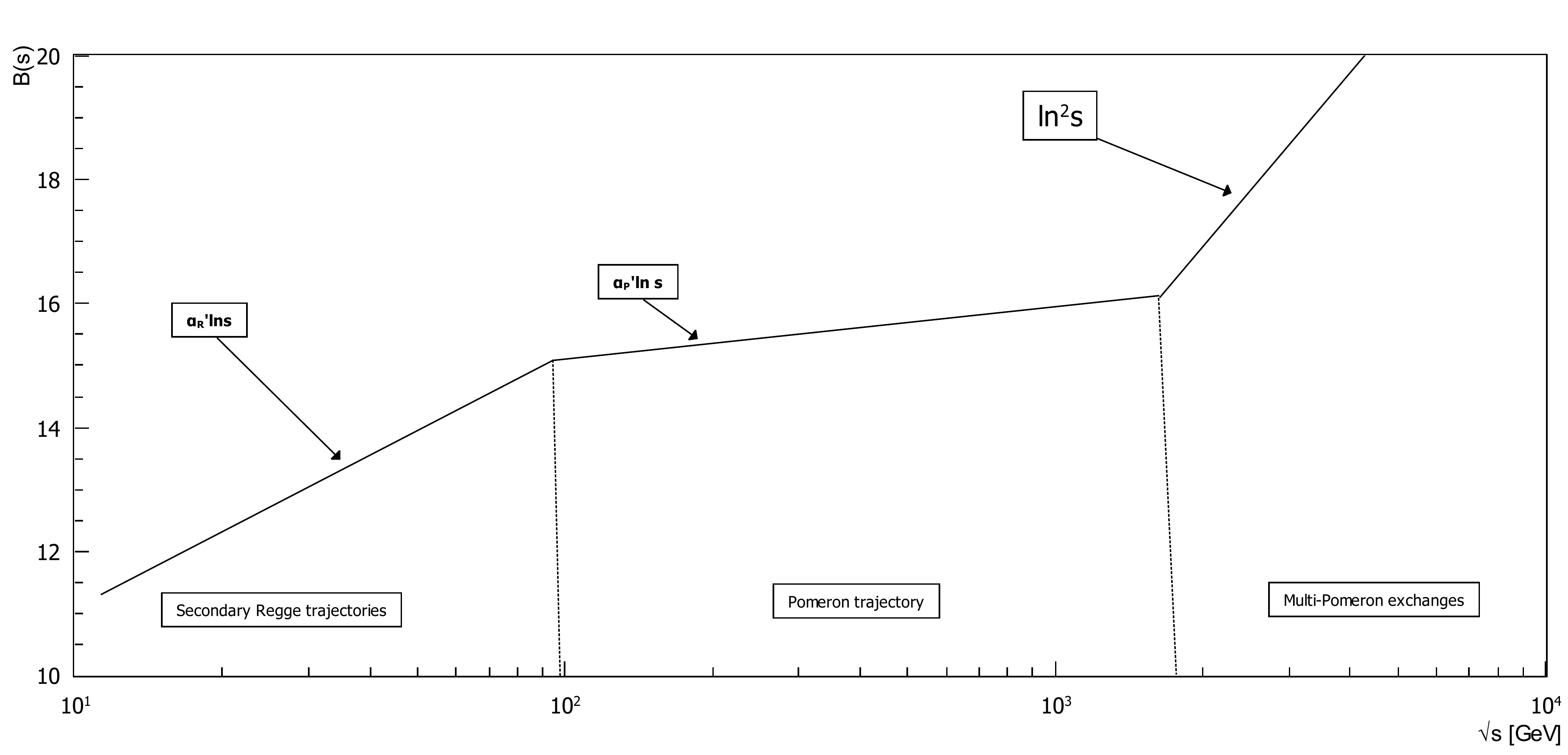}}
	\hspace{-1cm}
	\caption[ch2]{\small Three regimes of  the energy dependence of the diffraction cone slope parameter $B(s)$  in the  Regge-pole model amended with   contribution of multi--Pomeron exchanges  at the LHC energy range. {  Quantitative justification for the second break can be found
	in \cite{totem, atlas, tsr}.}}
\end{figure}
  The break in $B(s)$ {\it in the region of $\sqrt{s}\simeq 100$ GeV} should be treated as a signal of the  Pomeron dominance, but such { a} break has not  been observed { yet} (see \cite{tsr})\footnote{This conclusion  assumes an approximate equality of the diffraction cone slopes of $pp$ and $\bar p p$ elastic scattering in the energy range between CERN ISR and the LHC.}.

 { To this end, there is a problem regarding agreement of the whole set of data for the forward slope of the $pp$ elastic scattering differential cross--section with the Regge model predictions. The main point of this note is to make a  statement on  its existence.}
 
  The resolution of the Regge case might be  a twofold one. First, the energy range between CERN ISR and the LHC  should be thoroughly scanned experimentally with higher statistics without an assumption on the approximate equality of $pp$ and $\bar p p$ diffraction cone parameters. This  is correlated with the searches for possible signals of the Odderon contributions to the scattering amplitude (cf. \cite{mart,mart1} and references therein).  { Evidently, it would be interesting to perform experimental studies of the cross--sections difference $\Delta\sigma_{tot}(s)\equiv\sigma_{tot}^{\bar p p}(s)- \sigma_{tot}^{ p p}(s)$ in the unexplored energy region to extract a possible nonvanishing contribution of the Odderon at $t=0$.}
  
  Recent experimental  results on Odderon effects are based on the combined analysis of the D0 and TOTEM Collaborations results \cite{TD0}. { This analysis considers the LHC data set at $t=0$ together with Tevatron data  set in the dip region ($-t>0$) and assumes that  the same object (Odderon) gives contributions into the scattering amplitude in the both different regions of $t$. It neglects possible role of spin dependence also.}
  
  Another solution of  { the above} qualitative discrepancy is the use of  an approach supposing an alternative origin of the diffraction cone  slope $B(s)$ growth due to unitarization of an nput amplitude in the direct channel.  {  We further illustrate such possibility  using a specific form of unitarization.}
  
 \section{Unitarization and  { forward slope increase}} 
  The form of the input amplitude can be obtained using its  asymptotics in the region of large $l$--values. This { procedure }is quite similar to taking  Regge--form of the scattering amplitude. 
  
  To demonstrate this possibility we have considered a wide class of the geometrical models  \cite{epl,mpla}. { These models presuppose a factorized form of the input amplitude.}
The input amplitudes of these  models have  an energy independent  slopes of the diffraction cone and its unitarization generates the respective energy dependence. { The Regge--type models for the input amplitude do not allow clear separation of the unitarity effects since the logarithmic forward slope parameter increase is implied ab initio in the input amplitude.}

The slope associated with the interaction radius increases  at  low and moderate energy values, where the total cross--section does not demonstrate any increase  yet \cite{totihep}.  Growth of the { forward slope of } diffraction cone  is another reason for  unitarization  in geometrical approaches and this  serves as an argument for unitarization at low and medium energies. The results are found to be common for
different unitarization schemes. For definiteness we consider here the one based on the  $U$-matrix form of the scattering amplitude { (see for more details \cite{pl21} and references therein)}.
 
 The relation of the scattering amplitude $f(s,b)$\footnote{ Here the variable $b$ denotes a collision impact parameter.} with the input quantity $u(s,b)$  in the $U$--matrix approach (pure imaginary case, $f\to if$, $u\to iu$)\footnote{Note, that saturation of  the unitarity relation, i.e. when $\mbox{Im} f(s,b)\to 1$ at $s\to \infty$ and $b$--fixed, implies $\mbox{Re} f(s,b)\to 0$.}  in the impact parameter representation has a  rational form:
\begin{equation}\label{um}
f(s,b)=u(s,b)/[1+u(s,b)].
\end{equation}
It can easily be inverted 
\begin{equation}\label{umr}
u(s,b)=f(s,b)/[1-f(s,b)].
\end{equation}

Let us first consider the domain of fixed values of $s$ and large increasing impact parameters $b$. A factorized form should be valid in this limit for the function $u(s,b)$
since $$f(s,b)\simeq g(s)\exp(-\mu b)$$ and
\begin{equation}\label{uf}
u(s,b)/f(s,b)\to 1
\end{equation}
 in this region.   Assuming that such form for $u(s,b)$ takes place at finite impact parameter values, one  arrives to Eqs. (\ref{usb}) and (\ref{om}) that follow.

Then, to get an idea on the energy dependence of $u(s,b)$, we  evaluate energy dependence  at $s\to\infty$ and fixed $b$. Assuming saturation of the unitarity limit $f\to 1$ at $s\to\infty$, we obtain $u(s,b)\to\infty$. This assumption corresponds to the principle of maximal strength of strong interactions proposed long ago by Chew and Frautchi \cite{chew}.
Thus,  the limiting behavior $u(s,b)\to\infty$ at $s\to\infty$ and $b$--fixed corresponds aplication of the principle of maximal strength  in this approach. 

A general expression for the function $u(s,b)$ can be written as the ratio of elastic and inelastic distribution functions \cite{prev}:
\begin{equation}\label{prev}
u(s,b)={\sigma_{el}(s,b)}/{\sigma_{inel}(s,b)}.
\end{equation}
Eq. (\ref{prev}) characterizes relative spacial distribution of the elastic and inelastic  interactions.
Central impact parameter dependence of the function $u(s,b)$ implies central character of the elastic scattering and its domination over the inelastic interactions at small  and vanishing  ratio  ${\sigma_{el}(s,b)}/{\sigma_{inel}(s,b)}$ at large impact parameters. Conclusion on the central character of the elastic scattering is quite general and is not connected with a particular unitarization scheme used\cite{pl21}. 
The geometrical models use a  factorized form of the input amplitude, i.e. the function $u(s,b)$ is taken as a product:
\begin{equation}\label{usb}
u(s,b)=g(s)\omega(b),
\end{equation}
where dependence $g(s)\sim s^\lambda$ can  be associated with  the effective rate of the  kinetic energy to mass conversion \cite{carr,carr1,mcy}. This mass conversion rate reflects increasing  contribution with energy of the newly opened inelastic channels into the input amplitude $u(s,b)$. Its power-like dependence corresponds to the polynomial  boundedness of the function $u(s,b)$ which does not include elastic channels contributions \cite{echa}.

 As it was noted above, the function $\omega(b)$ is taken in the form to be consistent with the analyticity of the scattering amplitude in the Lehmann--Martin ellipse (see \cite{mt}) 
\begin{equation}\label{om}
\omega(b)\sim \exp(-\mu b).
\end{equation}
Of course, this conjecture is a simplest option for  the function $\omega(b)$. There are no reasons besides simplicity to expect that a function is to be  identical with its asymptotics.
The $u(s,b)$ can also be treated as a convolution of the two matter distributions $D_1\otimes D_2$ of the colliding hadrons having respective impact parameter dependencies  \cite{cy}.

 Unitarization adjusts energy behavior  of $B(s)$ to the experimental odservations. It has been shown that the resulting energy dependence of the diffraction cone slope  is consistent with the data and can be described  up to the LHC energies by a power--like preasymptotic function \cite{epl}.
 { It can be estimated that $\sqrt{s}\simeq 10^4$ GeV is the beginning of the transition energy region to asymptotics.}
 The  parameter $B(s)$ has  the following  asymptotic behavior  
 \begin{equation}\label{bes}
 B(s)\sim \ln^2 s
 \end{equation}
 which replaces  a power--like dependence  in the limit of $s\to\infty$. The   { transformation of these} two dependencies is  provided by the unitarization procedure \cite{mpla}.

Thus, unitarity plays an essential role at the LHC energies and the observed speeding up of the slope $B$ shrinkage can be qualitatively explained  as follows. The slope $B$ is related to the average   $$2B=\langle b^2 \rangle_{tot}\equiv {\int_0^\infty b^3db\sigma_{tot}(s,b)}/
{\int_0^\infty bdb\sigma_{tot}(s,b)}$$ and a major contribution to the interaction radius  $\langle b^2 \rangle_{tot}$ at the LHC energies is given by the inelastic interactions since the ratio $\sigma_{inel}/\sigma_{tot}\geq 2/3$.  Speed up of the $B$ increase at the LHC is a combined effect of a steady increase  of $\sigma_{inel}$ and reaching the maximal value of { the inelastic overlap function } $h_{inel}$ allowed by unitarity.  
Indeed, the inelastic overlap function $h_{inel}(s,b)$   is very close to its limiting value  
$h^{max}_{inel}=1/4$ in the  region of small and moderate impact parameters: $0\leq b\leq 0.4$ fm at the LHC energy $\sqrt{s}=13$ TeV \cite{alkin1,tsrg}. Deviation of $h_{inel}$ from its maximal value  is small and negative in this region of  impact parameters and  the following inequalities  take place:
\[
h_{el}>1/4>h_{inel},
\]
  where $h_{el}$ is the elastic overlap function and its deviation from $1/4$ is also small, but positive. { The unitarity relation in the impact parameter representation written in terms of the overlap functions $h_i(s,b)$ has the following form 
  	\[
  h_{tot}(s,b)= h_{el}(s,b)	+ h_{inel}(s,b).
  	\]
  }The unitarity stops growth of  the inelastic interaction intensity  in vicinity of $b=0$ and further increase of $\sigma_{inel}$,
  $$\sigma_{inel}=8\pi\int_0^\infty bdb h_{inel},$$
   becomes only possible due to proliferation of  the inelastic interactions into the region of higher  impact parameter values with the respective acceleration  of the slope parameter $B$ increase. 
  

There is also an 
enhancing effect connected with peripheral form of the inelastic overlap function developed with the energy growth. Therefore, 
one obtains an extra  augmentation of the speed  of $B$ growth due to contribution  of  emerging reflective scattering mode \cite{refl} (aka hollowness) associated with the hadron structure having a central core \cite{pl21}. 

\section{Conclusion}
The available experimental results indicate on { their}
inconsistency with the Regge model predictions. { We have discussed the Regge model problems with unobserved break in the energy dependence of the slope parameter in this regard.}   { This is one but not the only reason to address to the models of a different, e.g. geometrical origin}.  

{ The}  increasing behavior of the  diffraction cone { forward} slope with energy   is   { illustrated here} considering it   as  a result of  unitarization {of the input  explicit form of which is prescribed by the gometrical models} { when}  the { forward} slope  { of} the input amplitude   does not depend on energy.

 The  measurements of $B$ at the LHC  and their proposed  interpretations  make similar measurements in $pp$--scattering even at higher energies as well as at lower energies very promising for the studies of soft hadron interaction dynamics. 
\section*{Acknowledgements}
We are grateful to Evgen Martynov for a useful correspondence on the {   forward slope increase} 
 { and to the Reviewers for  many valuable comments.}

\end{document}